\documentclass[letterpaper, 10 pt, conference]{ieeeconf}  

\IEEEoverridecommandlockouts                              

\usepackage{graphicx}
\usepackage{float}
\usepackage{stfloats} 
\usepackage{caption}  

\usepackage{cite}

\usepackage{float}  
\usepackage{amsmath,amssymb,amsfonts}
\usepackage{algorithmic}
\usepackage{graphicx}
\usepackage{textcomp}
\usepackage{multirow}
\usepackage[table,xcdraw]{xcolor}
\usepackage{colortbl}
\def\BibTeX{{\rm B\kern-.05em{\sc i\kern-.025em b}\kern-.08em
    T\kern-.1667em\lower.7ex\hbox{E}\kern-.125emX}}

\makeatletter
\newcommand{\linebreakand}{%
  \end{@IEEEauthorhalign}
  \hfill\mbox{}\par
  \mbox{}\hfill\begin{@IEEEauthorhalign}
}
\makeatother

\title{\LARGE \bf
Human-Guided Feature Selection for Accurate Cardiomyocyte Dysfunction Classification
}

\author{Rana Raza Mehdi$^{1}$, Sukanya Sahoo$^{1}$, Sunder Neelakantan$^{1}$, Emilio A. Mendiola$^{1}$ \\ Kyle Myers$^{1}$, Sakthivel Sadayappan$^{2}$, Reza Avazmohammadi$^{1}$
\thanks{$^{1}$Texas A\&M University, College Station, TX, USA}
\thanks{$^{2}$University of Arizona, Tucson, AZ, USA}
}

\begin{document}

\maketitle
\thispagestyle{empty}
\pagestyle{empty}

\begin{abstract}

Early identification of cardiomyocyte dysfunction is a critical challenge for the prognosis of diastolic heart failure (DHF) exhibiting impaired left ventricular relaxation (ILVR). Myocardial relaxation relies strongly on efficient intracellular calcium (${\text{Ca}}^{2+}$) handling. During diastole, a sluggish removal of ${\text{Ca}}^{2+}$ from cardiomyocytes disrupts sarcomere relaxation, leading to ILVR \textit{at the organ level}. Characterizing myocardial relaxation \textit{at the cellular level} requires analyzing both sarcomere length (SL) transients and intracellular calcium kinetics (CK). However, due to the complexity and redundancy in SL and CK data, identifying the most informative features for accurate classification is challenging. To address this, we developed a robust feature selection pipeline involving statistical significance testing (p-values), hierarchical clustering, and feature importance evaluation using random forest (RF) classification to select the most informative features from SL and CK data. SL and CK transients were obtained from prior studies involving a transgenic phospho-ablated mouse model exhibiting ILVR (AAA mice) and wild-type as non-transgenic control mice (NTG). By iteratively refining the feature set, we trained a RF classifier using the selected reduced features. For comparison, we evaluated the performance of the classifier using the full set of original features as well as a dimensionally reduced set derived through principal component analysis (PCA). The confusion matrices demonstrated that the reduced feature set achieved comparable performance to the full feature set and outperformed the PCA-based approach, while offering better interpretability by retaining biologically relevant features. These findings suggest that a small, carefully chosen set of biological features can effectively detect early signs of cardiomyocyte dysfunction.
\newline

\indent \textit{\textcolor{black}{Clinical relevance}}\textcolor{black}{— The proposed feature selection approach facilitates detecting cardiomyocyte dysfunction at an earlier stage, offering clinicians precise, interpretable insights to support faster diagnosis and intervention decisions in conditions like diastolic dysfunction.}
\end{abstract}


\section{Introduction}
Diastolic heart failure (DHF) is characterized by impaired relaxation that significantly reduces the heart's ability to fill blood during diastole \cite{mehdi-2024}. The underlying cause often lies in cardiomyocyte impairment, which may develop due to abnormal calcium (${\text{Ca}}^{2+}$) cycling within cardiomyocytes \cite{Zile-2004, hamdani-2008}. A well-regulated Ca2+ cycle regulates contraction (${\text{Ca}}^{2+}$ influx) and relaxation (${\text{Ca}}^{2+}$ reuptake) \cite{neelakantan-2022abstract, neelakantan-2024abstract}. Disruptions in this cycle directly result in DHF, making the analysis of ${\text{Ca}}^{2+}$ cycling essential for evaluating cardiomyocyte impairment at the cellular level and potentially enhancing the diagnosis of DHF patients \cite{asp-2013}. Understanding the complex interaction between ${\text{Ca}}^{2+}$ handling and sarcomere length (SL) transients in cardiomyocyte function provides important insights into the relaxation procedure \cite{janssen-2010}.

\begin{figure}
            \centering
	            \includegraphics[scale=0.4]{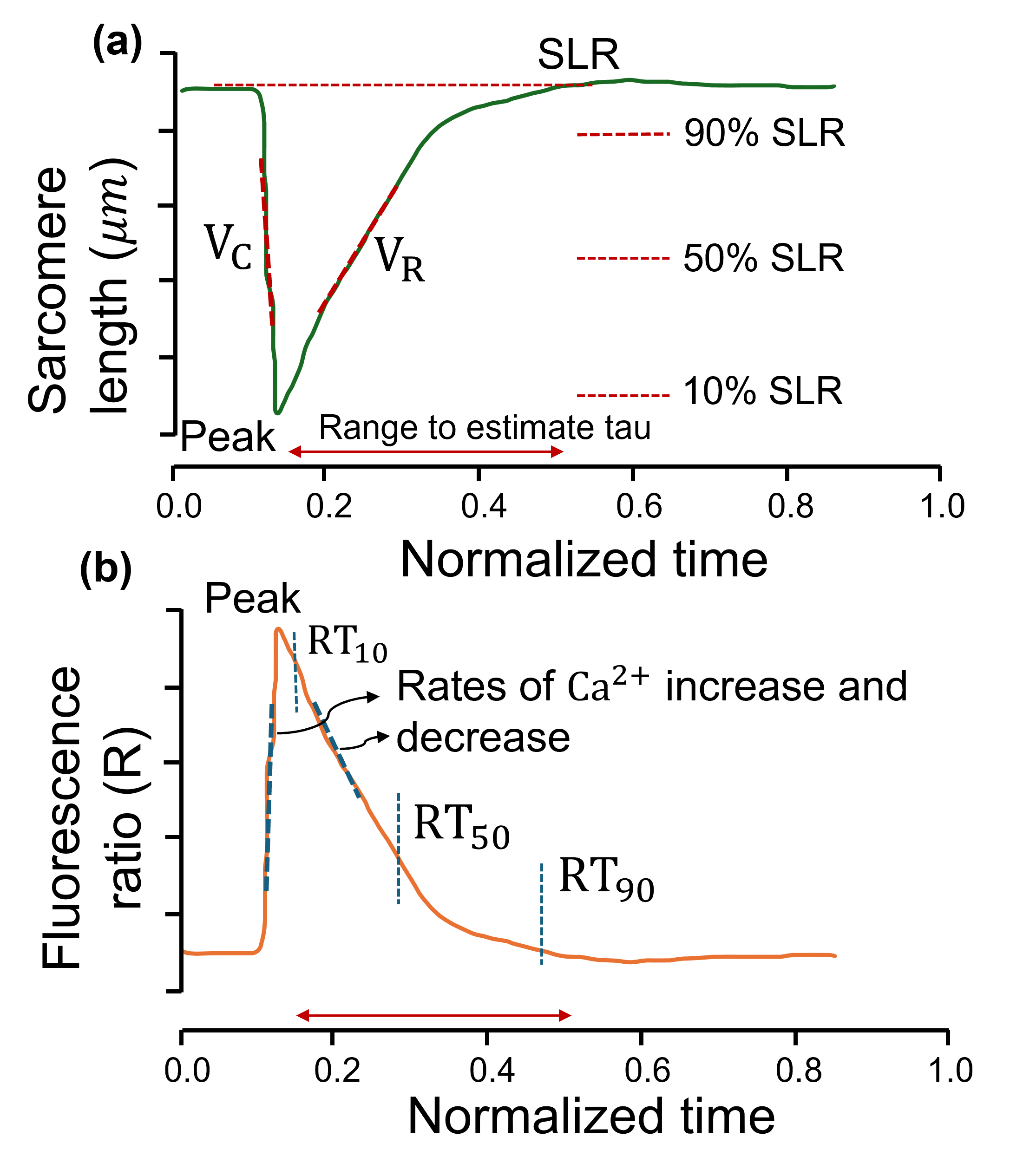}
	\caption{\textcolor{black}{The input features for preprocessing of principal component analysis and machine learning model from (a) sarcomere length transient and (b) calcium kinetics data. This figure is adopted from \cite{mehdi-2023}. Here, $V_C$ and $V_R$ represent contraction and relaxation velocities, SLR: sarcomere length at rest, tau: exponential decay constant, and RT: time to decay fluorescence ratio. More details are mentioned in the ``Feature extraction'' section.}}
	\label{features}
\end{figure}

\textcolor{black}{Despite the recognized importance of SL transients and intracellular calcium kinetics (CK) in understanding relaxation dysfunction, significant challenges remain to effectively identify key attributes for classification. This is because SL and CK datasets often contain too many overlapping and correlated features, and collecting a complete set of measurements is experimentally challenging and time-consuming. In addition, high dimensionality and missing values further complicate the analysis \cite{berisha-2021}. To address these limitations, we developed a feature selection pipeline that combines statistical testing, clustering of similar features, and feature importance ranking to reduce the number of features while preserving meaningful information. We then trained a random forest (RF) machine learning (ML) model on the selected features and compared its performance against models trained on principal component analysis (PCA)-reduced features and the full set of original features \cite{mehdi-2023RF}. We selected RF due to its strong performance on similar datasets, ability to handle small data with minimal tuning, and compatibility with feature importance analysis to support interpretable feature selection \cite{juhola-2018, williams-2020, mehdi-2023}}

\begin{figure*}[t]
            \centering
	            \includegraphics[scale=0.5]{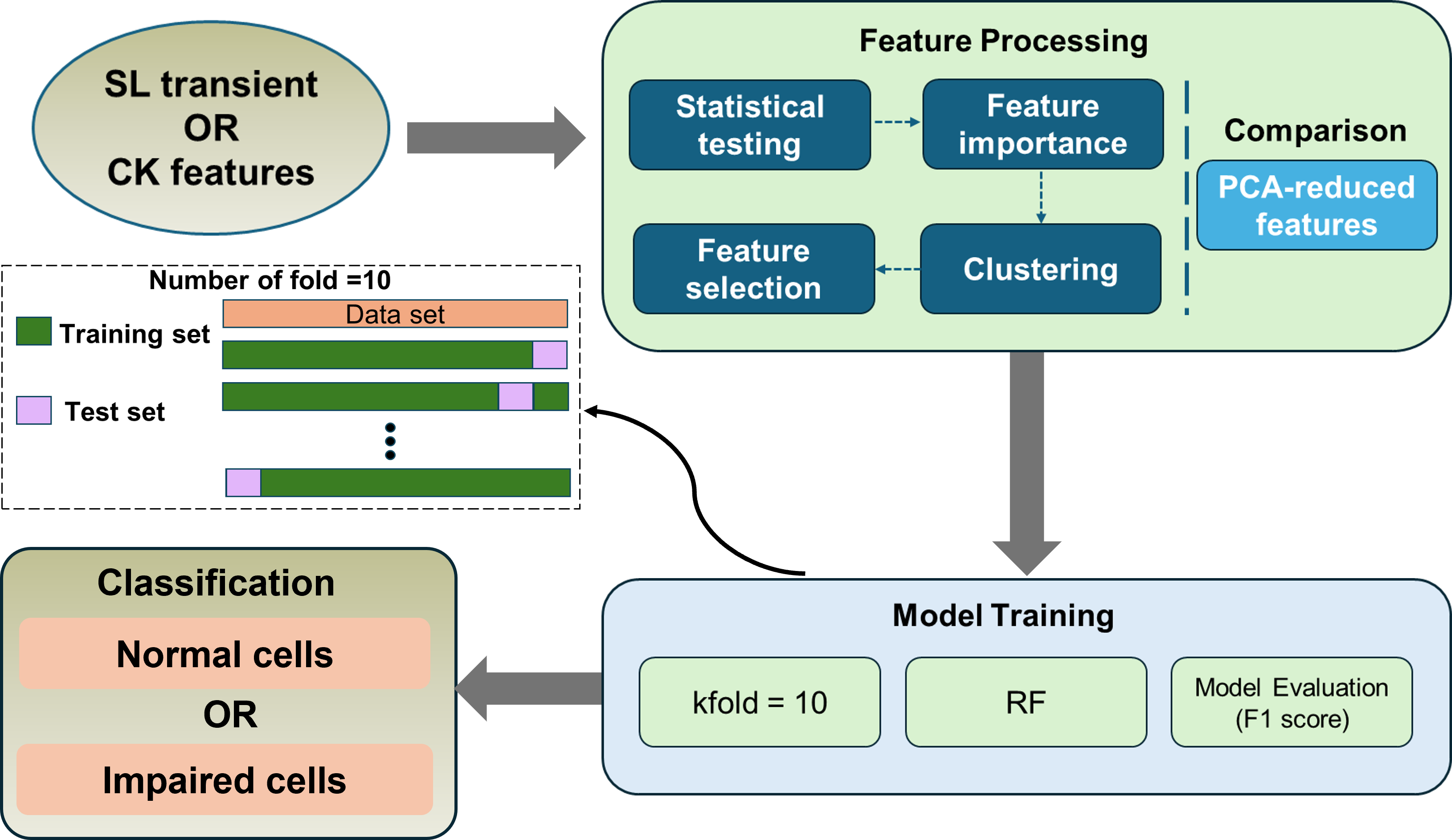}
	\caption{Flowchart summarizing feature selection and classification pipeline. Sarcomere length (SL) and calcium kinetics (CK) features were processed through a systematic selection pipeline involving statistical testing, clustering, and random forest-based feature importance. The selected features are compared to PCA-reduced features for evaluation. The selected features are then used to train a random forest classifier with 10-fold cross-validation, and performance is evaluated using the F1-score to classify normal and impaired cells.}
	\label{flowchart}
\end{figure*}

\section{METHODOLOGY}

\subsection{Murine models and ex-vivo data collection}
Murine models were employed to obtain cardiomyocytes with normal and impaired relaxation. Impaired cells were separated from a genetically modified phospho-ablated mouse model (AAA mice) by removing the phosphorylation sites Ser273, Ser282, and Ser302 on cardiac myosin binding protein-C. Normal cells were obtained from non-transgenic, wild-type mice (NTG). Ex-vivo data collection assessed SL transient and CK derived from cardiomyocytes of NTG and AAA mice. Specifically, SL transient data were collected from 126 cells (60 normal and 66 impaired), along with intracellular calcium transient assessments obtained from 116 cells (57 normal and 59 impaired). Cardiomyocytes were loaded with calcium-sensitive Fura-2 dye to gauge calcium transients and sarcomere contraction
estimated using a video-based detection system, mentioned in detail in \cite{mehdi-2023}.

\subsection{Feature extraction} \label{sec:features-ex}
SL and CK curves were used to obtain features to serve as inputs for the ML model. From SL curves, features included initial and contracted SLs, contraction and relaxation velocities, times for SL to decay and peak by specific percentages (10\%, 50\%, 90\%), fractional shortening, and decay constant (Fig. \ref{features}a). For CK curves, features included systolic and diastolic calcium levels, rates of calcium increase and decrease, decay constant, times for calcium decay and peak to specific percentages, and percentage change in calcium levels from diastole to systole (Fig. \ref{features}b).

\begin{table*}
\centering
{\large TABLE I}  
\caption*{Summary of sarcomere length (SL) and calcium kinetics (CK) data, including p-values, cluster assignments, and feature importance scores. Red-highlighted features were selected based on these values to reduce dimensionality and train the random forest classifier}
\vskip 0.2em
\label{feature-summary}
\begin{tabular}{l
>{\columncolor[HTML]{C0C0C0}}l 
>{\columncolor[HTML]{C0C0C0}}c 
>{\columncolor[HTML]{C0C0C0}}c 
>{\columncolor[HTML]{C0C0C0}}c l
>{\columncolor[HTML]{C0C0C0}}l 
>{\columncolor[HTML]{C0C0C0}}c 
>{\columncolor[HTML]{C0C0C0}}c 
>{\columncolor[HTML]{C0C0C0}}c }
                           & \multicolumn{4}{c}{\cellcolor[HTML]{FFDC96}\textbf{Sarcomere length}}                                                                                                                                                                                                                             &                     & \multicolumn{4}{c}{\cellcolor[HTML]{C8B3DE}\textbf{Calcium kinetics}}                                                                                                                                           \\
\multirow{-2}{*}{}         & \cellcolor[HTML]{FCEDDA}\textbf{Features}      & \multicolumn{1}{l}{\cellcolor[HTML]{FCEDDA}\textbf{Cluster}} & \multicolumn{1}{l}{\cellcolor[HTML]{FCEDDA}\textbf{p-value}} & \multicolumn{1}{l}{\cellcolor[HTML]{FCEDDA}\textbf{\begin{tabular}[c]{@{}l@{}}Feature \\ importance\end{tabular}}} &                     & \cellcolor[HTML]{E7E9FE}Features               & \cellcolor[HTML]{E7E9FE}Cluster & \cellcolor[HTML]{E7E9FE}p-value      & \cellcolor[HTML]{E7E9FE}\begin{tabular}[c]{@{}c@{}}Feature \\ importance\end{tabular} \\
\cellcolor[HTML]{D9FEFD}1  & Initial SL                                     & 4                                                            & 2.48e-03                                                     & 0.049286                                                                                                           &                     & \cellcolor[HTML]{FFCCC9}Diastolic Ca           & \cellcolor[HTML]{FFCCC9}4       & \cellcolor[HTML]{FFCCC9}5.757433e-11 & \cellcolor[HTML]{FFCCC9}0.213613                                                      \\
\cellcolor[HTML]{D9FEFD}2  & \cellcolor[HTML]{FFCCC9}Contracted SL          & \cellcolor[HTML]{FFCCC9}4                                    & \cellcolor[HTML]{FFCCC9}2.46e-03                             & \cellcolor[HTML]{FFCCC9}0.095874                                                                                   &                     & \cellcolor[HTML]{FFCCC9}Systolic Ca            & \cellcolor[HTML]{FFCCC9}1       & \cellcolor[HTML]{FFCCC9}2.254926e-01 & \cellcolor[HTML]{FFCCC9}0.039875                                                      \\
\cellcolor[HTML]{D9FEFD}3  & \cellcolor[HTML]{FFCCC9}Contraction velocity   & \cellcolor[HTML]{FFCCC9}2                                    & \cellcolor[HTML]{FFCCC9}4.98e-06                             & \cellcolor[HTML]{FFCCC9}0.114271                                                                                   &                     & Ca increase rate                               & 1                               & 2.529702e-01                         & 0.044801                                                                              \\
\cellcolor[HTML]{D9FEFD}4  & \cellcolor[HTML]{FFCCC9}Relaxation velocity    & \cellcolor[HTML]{FFCCC9}5                                    & \cellcolor[HTML]{FFCCC9}3.84e-09                             & \cellcolor[HTML]{FFCCC9}0.162466                                                                                   &                     & Ca decrease rate                               & 6                               & 2.786914e-01                         & 0.034543                                                                              \\
\cellcolor[HTML]{D9FEFD}5  & \cellcolor[HTML]{FFCCC9}Time for SL 10\% decay & \cellcolor[HTML]{FFCCC9}1                                    & \cellcolor[HTML]{FFCCC9}6.74e-03                             & \cellcolor[HTML]{FFCCC9}0.057725                                                                                   &                     & Time for Ca decay 10\%                         & 3                               & 2.455269e-04                         & 0.086792                                                                              \\
\cellcolor[HTML]{D9FEFD}6  & Time for SL 50\% decay                         & 1                                                            & 1.12e-02                                                     & 0.061710                                                                                                           &                     & \cellcolor[HTML]{FFCCC9}Time for Ca decay 50\% & \cellcolor[HTML]{FFCCC9}3       & \cellcolor[HTML]{FFCCC9}2.246343e-06 & \cellcolor[HTML]{FFCCC9}0.201821                                                      \\
\cellcolor[HTML]{D9FEFD}7  & Time for SL 90\% decay                         & 1                                                            & 1.59e-02                                                     & 0.057015                                                                                                           &                     & Time for Ca decay 90\%                         & 3                               & 5.532377e-03                         & 0.077900                                                                              \\
\cellcolor[HTML]{D9FEFD}8  & \cellcolor[HTML]{FFCCC9}Time for SL 10\% peak  & \cellcolor[HTML]{FFCCC9}3                                    & \cellcolor[HTML]{FFCCC9}8.13e-03                             & \cellcolor[HTML]{FFCCC9}0.065552                                                                                   &                     & Time for Ca peak 10\%                          & 5                               & 6.727202e-01                         & 0.025998                                                                              \\
\cellcolor[HTML]{D9FEFD}9  & Time for SL 50\% peak                          & 2                                                            & 7.88e-04                                                     & 0.051430                                                                                                           &                     & \cellcolor[HTML]{FFCCC9}Time for Ca peak 50\%  & \cellcolor[HTML]{FFCCC9}5       & \cellcolor[HTML]{FFCCC9}2.790539e-03 & \cellcolor[HTML]{FFCCC9}0.050407                                                      \\
\cellcolor[HTML]{D9FEFD}10 & Time for SL 90\% peak                          & 1                                                            & 2.14e-02                                                     & 0.053965                                                                                                           &                     & Time for Ca peak 90\%                          & 5                               & 3.624180e-01                         & 0.035852                                                                              \\
\cellcolor[HTML]{D9FEFD}11 & Fractional shortening                          & 5                                                            & 4.07e-08                                                     & 0.160933                                                                                                           &                     & \cellcolor[HTML]{FFCCC9}Percent change of Ca   & \cellcolor[HTML]{FFCCC9}2       & \cellcolor[HTML]{FFCCC9}9.756891e-03 & \cellcolor[HTML]{FFCCC9}0.104323                                                      \\
\cellcolor[HTML]{D9FEFD}12 & Decay constant                                 & 1                                                            & 6.74e-01                                                     & 0.069774                                                                                                           & \multirow{-14}{*}{} & Decay constant                                 & 4                               & 1.137651e-03                         & 0.084074                                                                             
\end{tabular}
\end{table*}

\begin{figure*}[b]
    \centering
    \includegraphics[width=0.8\linewidth]{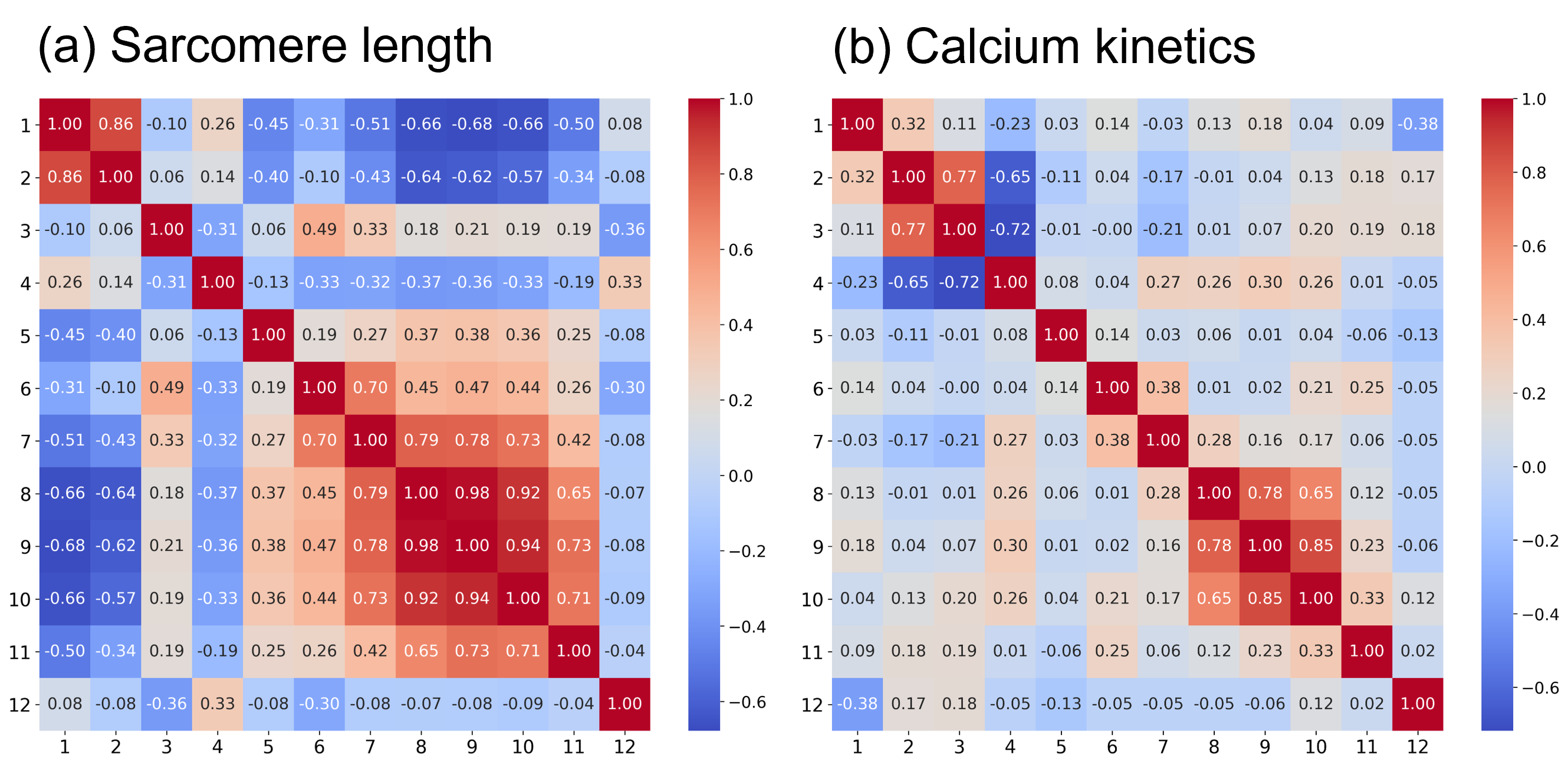}
    \caption{Correlation heatmaps for (a) sarcomere length (SL) and (b) calcium kinetics (CK) features. The heatmaps visualized the pairwise correlations among features, where hierarchical clustering was used to group highly correlated features, thereby reducing redundancy by grouping similar ones.}
    \label{heatmap}
\end{figure*}

\subsection{Feature selection pipeline}
While the dataset contains only 12 features, practical challenges arise from the difficulty of collecting all features simultaneously. Therefore, identifying a reduced set of the most relevant features is important to build a robust classification model without compromising on its performance. We developed a structured feature selection pipeline as follows:  

\textbf{1. Statistical significance testing:} Features were first ranked based on their ability to distinguish between NTG and AAA cells using p-values derived from a two-sample t-test.  

\textbf{2. Clustering of correlated features:} Hierarchical clustering was performed to group highly correlated features, reducing redundancy and ensure diversity. 

\textbf{3. Feature importance ranking:} RF classifier was used to rank feature importance within each cluster.

\textbf{4. Final selection:} The most significant feature was selected from each cluster based on the combined p-values and feature importance. This resulted in a reduced feature set for training the classifier.

This feature engineering process enabled the training of an effective classifier with a minimal set of features, addressing practical constraints while achieving performance comparable to using the full set of features.

\subsection{Principal component analysis and random forest classification}
We identified the uncorrelated attributes by applying PCA to the original dataset. PCA captured the maximum variance in the dataset by transforming the original correlated attributes into a set of uncorrelated principal components (PCs), ranked in order of the amount of variance. We retained PCs explaining 95\% of the total variance and used them as inputs for the RF classifier. The performance of the RF model trained with PCA-reduced features was compared with the model trained using the selected reduced feature set.

\subsection{Random Forest classifier with cross-validation}
RF classifiers were trained separately on the reduced feature set, PCA-transformed feature set, and the full original feature set to distinguish normal (NTG) from the impaired cardiomyocytes (AAA). For validation, we used 10-fold cross-validation, where the dataset was split into 10 equal-sized subsets. Each subset was used as a testing set once, while the remaining subsets served as the training set. This was repeated 10 times and evaluated using accuracy, precision, recall and F1-score and visualized correctly and incorrectly classified samples for NTG and AAA cells using confusion matrices.

\section{RESULTS}
\subsection{Selected features and their relevance}
Table I presents the chosen features from both SL and CK data, along with their corresponding p-values, cluster assignments, and feature importance scores. Features marked in red were identified via a systematic selection strategy that combined statistical significance testing, hierarchical clustering, and RF-driven importance ranking. This approach successfully reduced the feature set while preserving the critical features required to differentiate between normal and impaired cells.

\subsection{Feature correlation and clustering}
Fig. \ref{heatmap} displays correlation heatmaps illustrating the relationships between features for both SL and CK data. To reduce redundancy while preserving a variety of complementary feature representations, hierarchical clustering was employed to cluster features with high correlation.

\subsection{Classification performance using different feature sets}
We evaluated the classification performance of RF classifiers using three feature sets: the selected features from our proposed method, PCA-reduced features, and the complete set of original features. To simplify the results, the confusion matrix in Fig. \ref{confusion-matrix} combined the SL and CK datasets into a single representation. The suggested approach achieved high classification performance (F1-score: 0.846), comparable to the outcome with the complete feature set, while surpassing the PCA-based method (F1-score: 0.769). As illustrated in Table II, the RF classifier, trained on the selected features, yielded the highest F1-score, highlighting the effectiveness of our approach in striking a balance between performance and feature efficiency.

\section{Discussion}
\subsection{Feature selection through domain-specific knowledge}
The suggested feature selection process demonstrated that integrating domain-specific knowledge, statistical evaluations, clustering, and RF-based importance rankings can efficiently decrease the feature set while maintaining classification performance. By focusing on features that hold crucial biological significance, we effectively reduced redundancy and retained complementary information from both SL and CK datasets (Table I). This approach is particularly advantageous in the biological experimental field, where collecting all features simultaneously may be impractical due to experimental limitations. Unlike purely data-driven dimensionality reduction techniques, such as PCA, our method retained interpretability by keeping individual features directly associated with key biological processes.

\begin{figure*}[t]
    \centering
    \includegraphics[width=1.0\linewidth]{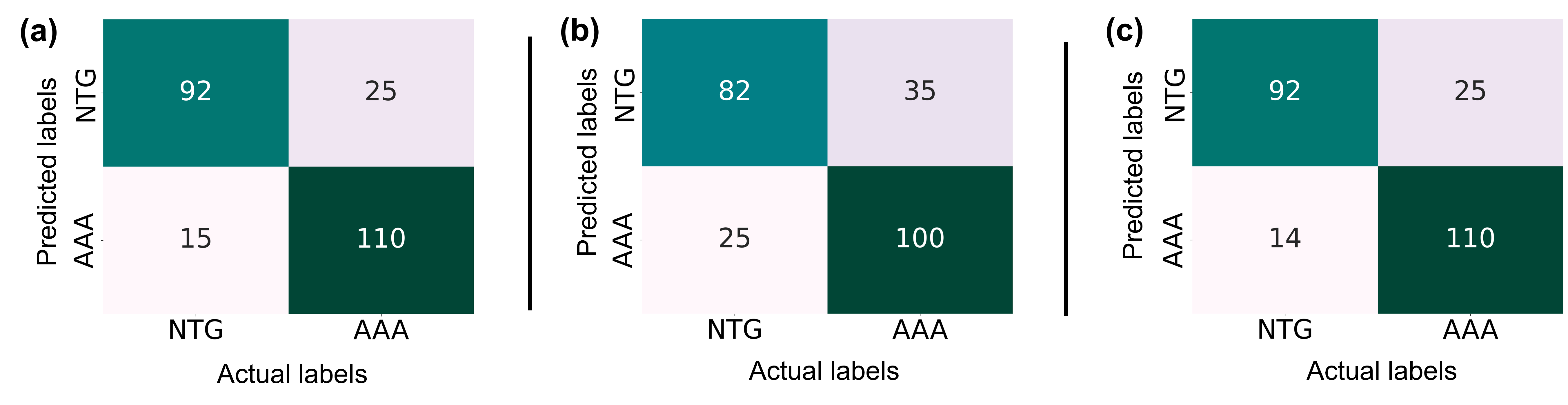}
    \caption{Confusion matrices for the classification of normal (NTG) and impaired (AAA) cells using (a) the proposed feature selection method, (b) PCA-reduced features, and (c) all original features. The proposed method demonstrated comparable performance to using all original features while outperforming the PCA-based approach.}
    \label{confusion-matrix}
\end{figure*}

\subsection{Performance comparison of selected features and PCA-based reduction}
The results showed that the RF classifier trained on the selected features achieved comparable or better classification performance than using all original features and significantly outperformed the PCA-based reduction (Fig. \ref{confusion-matrix} and Table II). The confusion matrices and F1-scores highlighted the effectiveness of the proposed method in correctly classifying normal and impaired cells. Although PCA is commonly utilized to reduce dimensionality, it often undermines interpretability by converting original features into PCs. In contrast, our approach to identify features maintained distinct feature identities, aiding in a clearer understanding of the importance of specific features in predicting cardiomyocyte dysfunction.

\begin{table}[H]
\centering
{\large TABLE II}  
\caption*{Performance comparison of classification methodologies using accuracy, precision, recall, and F1-score.}
\vskip 0.2em
\label{f1-scores}
\begin{tabular}{llllc}
\hline
Methodology     & Accuracy & Precision & Recall & F1-score \\ \hline
Proposed method & 0.835    & 0.815     & 0.88   & 0.846    \\
PCA             & 0.752    & 0.741     & 0.80   & 0.769    \\
All features    & 0.835    & 0.815     & 0.88   & 0.846    \\ \hline
\end{tabular}
\end{table}

\subsection{Biological relevance and limitations}
\textcolor{black}{The selected features offer significant biological insights into the processes that facilitate the dysfunction of cardiomyocyte relaxation function (Table I). Key selected features, such as contraction and relaxation velocities, diastolic calcium levels, and time-dependent SL and CK changes, align with known biomarkers of impaired calcium handling and sarcomere dysfunction. This suggests that these features not only serve as effective predictors of dysfunction but also reflect critical aspects of the underlying pathology. However, the study's dependence on murine models and a relatively small dataset may limit the generalizability of the findings to humans. Future work will be focused on validating these selected features in broader, diverse data sets and exploring their clinical importance further.}

\section*{ACKNOWLEDGMENT}

This work was supported by the National Institutes of Health (R00HL138288) and the National Science Foundation (2244995) to R.A. and the American Heart Association (25PRE1377496) to R.R.M.

\bibliography{Refs}

\end{document}